\documentclass[fleqn,usenatbib]{mnras}

\usepackage{newtxtext,newtxmath}

\usepackage[T1]{fontenc}
\usepackage{ae,aecompl}

\usepackage{graphicx}	
\usepackage{amsmath}	
\usepackage{caption}
\usepackage{pdflscape}
\usepackage{subcaption}
\usepackage{siunitx}
\usepackage[normalem]{ulem}

\title[Gas dispersal from Class III circumstellar disc]{Rapid CO gas dispersal from NO~Lup's class~III circumstellar disc}

\author[J. B. Lovell et al.]{J. B. Lovell$^{1}$\thanks{E-mail: jl638@cam.ac.uk}, G. M. Kennedy$^{2,3}$, S. Marino$^{1}$, M. C. Wyatt$^{1}$, M. Ansdell$^4$, M. Kama$^{5,1}$, \newauthor C. F. Manara$^6$, L. Matr\`a$^7$, G. Rosotti$^8$, M. Tazzari$^1$, L. Testi$^{6,9}$, J. P. Williams$^{10}$ \\
$^1$Institute of Astronomy, University of Cambridge, Madingley Road, Cambridge, CB3 0HA, UK\\
$^2$Department of Physics, University of Warwick, Coventry, CV4 7AL, UK \\
$^3$Centre for Exoplanets and Habitability, University of Warwick, Gibbet Hill Road, Coventry CV4 7AL, UK\\
$^4$National Aeronautics and Space Administration Headquarters, 300 E Street SW, Washington DC 20546, USA\\
$^5$Tartu Observatory, University of Tartu, 61602 T\H{o}ravere, Estonia \\
$^6$European Southern Observatory, Karl-Schwarzschild-Strasse 2, 85748, Garching bei M{\"u}nchen, Germany\\
$^7$School of Physics, National University of Ireland Galway, University Road, Galway, Ireland\\
$^8$Leiden Observatory, Leiden University, P.O. Box 9513, NL-2300 RA Leiden, the Netherlands\\
$^{9}$INAF - Osservatorio Astrofisico di Arcetri, L.go E. Fermi 5, 50125, Firenze, Italy\\
$^{10}$Institute for Astronomy, University of Hawai'i at Mänoa, Honolulu, HI, USA\\}

\date{Accepted 2020 November 17. Received 2020 November 17; in original form 2020 October 16}

\pubyear{2020}

\begin{document}
\label{firstpage}
\pagerange{\pageref{firstpage}--\pageref{lastpage}}
\maketitle

\begin{abstract}
We observed the K7 class~III star NO~Lup in an ALMA survey of the 1-3\,Myr Lupus association and detected circumstellar dust and CO gas. 
Here we show that the J = 3-2 CO emission is both spectrally and spatially resolved, with a broad velocity width ${\sim}19$\,km\,s$^{-1}$ for its resolved size ${\sim}1\arcsec$ (${\sim}130$\,au). 
We model the gas emission as a Keplerian disc, finding consistency, but only with a central mass of ${\sim}11M_{\odot}$, which is implausible given its spectral type and X-Shooter spectrum. 
A good fit to the data can also be found by modelling the CO emission as outflowing gas with a radial velocity ${\sim}22$\,km\,s$^{-1}$. 
We interpret NO~Lup's CO emission as the first imaged class~III circumstellar disc with outflowing gas. 
We conclude that the CO is continually replenished, but cannot say if this is from the break-up of icy planetesimals or from the last remnants of the protoplanetary disc. 
We suggest further work to explore the origin of this CO, and its higher than expected velocity in comparison to photoevaporative models.
\end{abstract}

\begin{keywords}
circumstellar matter - planetary systems - submillimetre: planetary systems.
\end{keywords}


\section{Introduction}
\label{sec:intro}
Stars are born with protoplanetary discs containing large quantities of primordial gas and dust which persist for several Myr before dispersing on rapid ${\sim}$0.1\,Myr timescales \citep{Ercolano17}. 
Circumstellar discs are also seen around older stars (${\gtrapprox}$10\,Myr), known as debris discs, where the dust and gas is inferred to be secondary, created in the destruction of planetesimals that must be volatile-rich to replenish the gas \citep{Wyatt08, Dent14, Marino16, Moor17, Matra17, Kral17}. 
The transition between the two types of disc is not well understood, but is thought to involve a combination of gas being accreted onto the star, or being expelled from the system by disc winds driven by photoevaporation or magneto-hydrodynamics (MHD), as well as planet formation processes \citep{Williams11, Wyatt15, Lesur20}. 
Class~III stars are those in star forming regions for which infrared emission shows an absence of hot dust suggesting that the star’s protoplanetary disc has either recently dispersed or is in the process of dispersal \citep{Adams87}. 
However, these stars usually have limited constraints on the presence of cold dust or gas with which to constrain their nature. \\
\newline
A recent ALMA survey of class~III stars in Lupus found dust in several systems with dust masses orders of magnitude lower than protoplanetary disc levels and consistent with originating in the progenitors of debris discs seen at later ages, though this does not rule out the possibility that this dust is a remnant of the protoplanetary disc \citep{Lovell20}. 
For one of these class~III stars, NO~Lup, J = 3-2 CO gas was also detected, allowing a more thorough assessment of the nature of its circumstellar environment. 
Comparison of the inferred mass of CO with that of the 10\,Myr old M star TWA~7, for which the level was consistent with secondary production \citep[i.e., $0.8{-}80{\times}10^{-6}M_\oplus$, see][]{Matra19B}, shows that the CO could indeed be secondary, though as for the dust interpretation, such plausibility does not preclude the possibility that the CO is primordial. \\
\newline
This paper presents a detailed analysis of the CO detected towards NO~Lup, using spectral and spatial data not reported in \citet{Lovell20}, constraining the kinematic structure of the gas disc. 
While circumstellar gas is usually seen in a Keplerian disc, other gas morphologies are possible, such as an outflowing component which could be a feature of either a depleting primordial disc \citep{Haworth20} or a debris disc. 
In $\S$\ref{sec:context} we provide a summary of the NO~Lup system, and introduce new X-Shooter observations. 
In $\S$\ref{sec:furtheranalysis} we extend the analysis of the CO gas by showing the problem with modelling this as a Keplerian disc, and show better consistency in $\S$\ref{sec:modelling2} by modelling this with an outward radial velocity. 
We interpret our results in $\S$\ref{sec:discussion}, and conclude in $\S$\ref{sec:conclusions}.

\section{NO~Lup in Context}
\label{sec:context}
\begin{figure}
    \includegraphics[width=1.0\columnwidth]{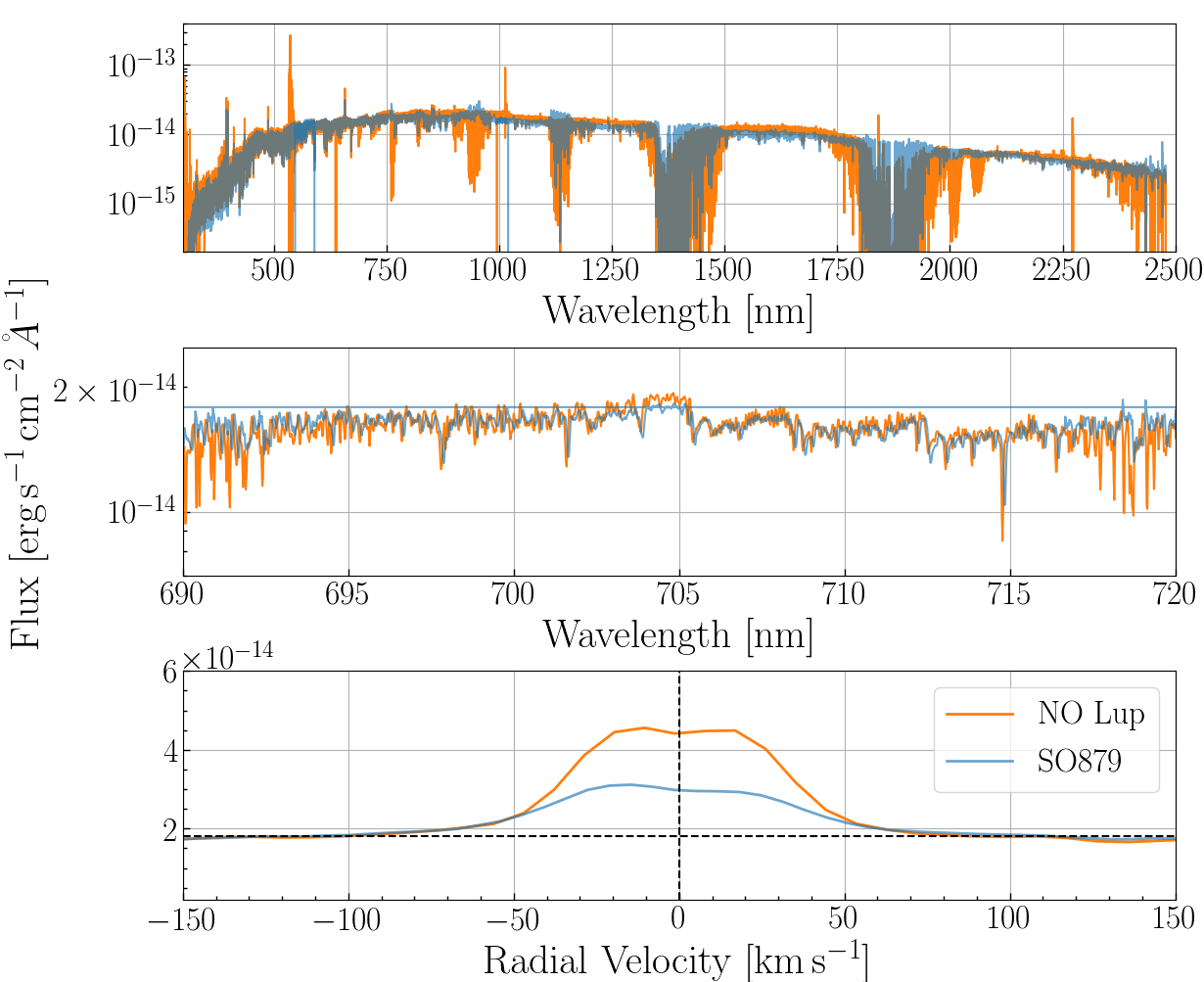}
    \caption{X-Shooter spectra for NO~Lup and SO879. Top: full X-Shooter range. Middle: zoomed-in 690-720\,nm range. Bottom: H${\alpha}$ line. SO879's flux is scaled to NO~Lup's median source flux between 698-702\,nm, and is wavelength corrected to have a common RV with NO~Lup based on SO879's DR2 RV of $30.10\pm0.21$\,km\,s$^{-1}$ \citep{Gaia18}.}
    \label{fig:XShoot}
\end{figure}

NO~Lup (2MASS~J160311.8-323920) is located in Cloud I of Lupus at ${\alpha}{=}$16:03:11.812, ${\delta}{=}$-32:39:20.31 (J2000) at a distance of $133.7{\pm}0.7$\,pc \citep{Gaia18}. 
NO~Lup has previously been classified spectrally as non-accreting \citep{Cieza13}, and as a K7 star \citep{Krautter97}. 
The latter is consistent with its Gaia DR2 temperature, $T_{\rm{eff}}{=}$3994$K$, and stellar luminosity, $L_\star{=}0.287L_\odot$ 
and its well constrained spectral energy distribution \citep[given observations with \textit{WISE}, \textit{Spitzer}, and \textit{2MASS}, which found the blackbody planetesimal belt radius as $R_{\rm{BB}}{=}3.2{\pm}0.3$\,au, see][]{Lovell20}. 
Analysing the X-Shooter spectrum of NO~Lup (project 093.C-0506 A), we confirm both spectral and accretion analyses. 
By reducing this data using the standard pipeline version 3.5.0 \citep{Modigliani10}, we show in Fig.~\ref{fig:XShoot} (top and middle) the full and zoomed-in X-Shooter spectra for NO~Lup, which are consistent with the spectral features of the well characterised class~III K7 star, SO879. 
In the lower panel we show the H${\alpha}$ emission line with an EW=$-2.80{\pm}0.15{\si{\angstrom}}$, which we find to be centred on a radial velocity (RV) of $-3.5{\pm}2.0$\,km\,s$^{-1}$, consistent with the Gaia DR2 RV of $-1.93{\pm}4.08$\,km\,s$^{-1}$, and the average RV of Lupus stars, $\rm{RV_{Lup}}=2.8{\pm}4.2\rm{km\,s^{-1}}$ \citep{Frasca17}. 
This EW is consistent with non-accretion and only slightly higher than, but also consistent with, the line width of non-accreting SO879. 
\citet{Hardy15} estimated the stellar mass of NO~Lup to be $M_{\star}{=}0.7M_{\odot}$. 
With the models of \citet{Siess00} and \citet{Baraffe15}, and the Gaia DR2 $L_\star$ and T$_{\rm{eff}}$, we estimated $M_\star$ between 0.7-0.8$M_\odot$, consistent with the literature. 
Although NO~Lup has significant emission above the photosphere at 12 and 24$\mu$m, these excesses are small, resulting in mid-IR spectral slopes steeper than those of protoplanetary discs. 
Unresolved continuum emission was detected by \textit{ALMA}, implying a dust mass of $0.036{\pm}0.007M_{\oplus}$ and disc radius <56\,au \citep{Lovell20}. 
These observations also detect CO J=3-2 line emission, with $F_{\rm{CO}}{=}0.29{\pm}0.07$\,Jy\,km\,s$^{-1}$, and a width of ${\sim}19$\,km\,s$^{-1}$ (between -11.0 to +8.1 $\rm{km} \, \rm{s}^{-1}$), consistent with being centred on the stellar RV discussed above. 
Assuming this emission is optically thin and in LTE at $T{=}50K$, this flux corresponds to a CO gas mass of $4.9{\pm}1.1{\times}10^{-5}M_{\oplus}$, which for an ISM CO abundance implies a gas-to-dust ratio of $1.0{\pm}0.4$ \citep{Lovell20}. We note that a wider temperature range of $20{-}100$\,K, consistent with the range of literature gas temperatures, could change this gas mass by at most a factor of ${\sim}$2 \citep[see equations 2 and 8 of][]{Matra17}. 

\begin{figure*}
    \includegraphics[width=1.0\textwidth]{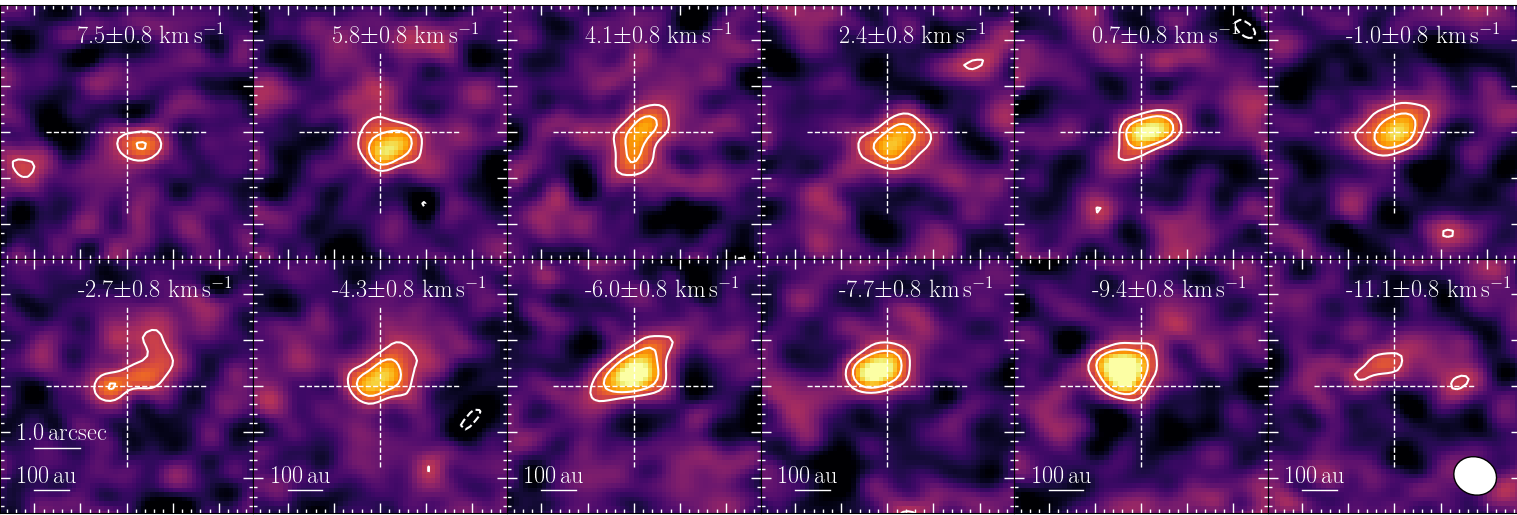}
    \caption{Averaged channel map (4 ${\sim}$0.4\,km\,s$^{-1}$ channels per plot), showing the $\pm3$ and $5\sigma$ emission significance. In all plots North is up, East is left, contour lines represent $\pm3$ and $\pm5\sigma$, 100\,au scale bars are shown in the lower-left of the bottom panel plots, the synthesised beam is shown in the bottom-right plot, and the major axis ticks are $1''$ wide. }
    \label{fig:chanMapNOLup}
\end{figure*}

\section{The Problem with a Keplerian Disc}
\label{sec:furtheranalysis}
Assuming the gas is in Keplerian rotation, the width of the CO line can be used to estimate the spatial extent of the emission. 
For example, if the star has a mass ${\sim}0.7M_{\odot}$ (see $\S$\ref{sec:context}), the line width indicates a radius of ${\sim}10$\,au (or smaller if the disc is not edge on), well below the spatial resolution limit of our measurements. 
However, we find that the CO emission is spatially resolved. 
Fig.~\ref{fig:chanMapNOLup} shows that the centroid of the CO emission transitions from the South-West to the North-East when binned in channels with decreasing RV, as expected for emission that originates in a disc inclined to the plane of the sky. 
To resolve this emission given a beam of ($0.94\times0.82$)$\arcsec$ (at position angle $70^{\circ}$), the disc emission would have to have a radius ${\gtrapprox}\,50$\,$\rm{au}$. 
This however would then be inconsistent with the previously inferred ${\sim}10$\,au radius for the assumed stellar mass. 
This conclusion can also be illustrated with the position-velocity (PV) diagram in Fig.~\ref{fig:NOLupPVDiag} (see left-hand plot 1), produced with a 36$^{\circ}$ position angle and 2$\arcsec$ slit width, which shows how the offset varies with radial velocity. 
This agrees with the previous analysis that the velocities are unexpectedly high at a large separation (nearly $10\,\rm{km}\,\rm{s}^{-1}$ at ${\sim}\,0.5\arcsec{=}66$\,au). 
The curves and radial extension line on this same plot show that for gas to be in a Keplerian orbit the stellar mass would have to be ${\sim}11M_{\odot}$, i.e., much higher than $0.7M_{\odot}$. \\
\newline
We use a simple model to explore this Keplerian disc interpretation by assuming the gas is in a Keplerian orbit with the stellar mass of NO~Lup left as a free parameter, $M_\star$. 
This \textit{Disc Model} computes the density of emitting CO for each pixel in a cube that has RA, Dec, and line of sight, $z$, as axes. 
The orientation, position and RV of the disc in the cube are set by the ascending node, $\Omega$ (i.e., the PA of the disc major axis), the inclination, $i$, a phase centre offset in RA and Dec (x$_{\rm{off}}$ and y$_{\rm{off}}$), the systemic velocity, $v_{\rm{sys}}$, and the fixed distance to NO~Lup of 133.7\,pc (see $\S$\ref{sec:context}). 
The model assumes that the CO density can be modelled with a total flux, $F$, and a power-law radial profile, defined between $r_{\rm{in}}$ and $r_{\rm{out}}$ with index $p$, for which the volume density goes as $r^{p-1}$. 
The model assumes a Gaussian scale height with a fixed aspect ratio, $\sigma_{h}=0.05$, meaning that the surface density goes as $\Sigma {\propto} r^p$. 
We fitted the visibilities of channels -15.1 to ${+}6.9$\,km\,s$^{-1}$ from the CO line \citep[as done in][]{Kennedy19}, using the \textit{emcee} package \citep{FM13}, with 40 walkers and 2000 steps to achieve convergence. 
The fit gives reasonable results, showing no residuals in the PV diagram ${>3}\sigma$ (see Fig.~\ref{fig:NOLupPVDiag}, plots 2 and 3), with the best-fit model values shown in Table~\ref{tab:BFMod}. 
We note that the best-fit $v_{\rm{sys}}{=}{-3.8}$\,km\,s$^{-1}$ is consistent with that in $\S$\ref{sec:context}, and that this model did not constrain $p$, which uniformly varied between the imposed limits of -2 to 1. 
A similar result was found with a Gaussian radial distribution. 
This fitting procedure therefore comes to the same conclusion as shown earlier in requiring a stellar mass $M_{\star}{=}11.1{\pm}1.9M_{\odot}$ that is significantly greater than expected. 
Since the stellar luminosity rules out ${\sim}10$ K-type stars within 10s of au of NO~Lup, and the X-Shooter spectra rules out NO~Lup being a mis-classified star, we conclude that we cannot consistently interpret the CO emission as originating from a Keplerian disc.

\begin{table}
    \centering
    \caption{Best-fit model parameters. D: \textit{Disc Model}, and O: \textit{Outflow Model}.}
    \begin{tabular}{c|c|c|c|c|c|c}
         \hline
         \hline
         Mod & $\Omega$ & $i$ & $r_{\rm{in}}$ & $r_{\rm{out}}$ & $M_\star$ & $v_r$ \\
         & deg & deg & au & au & $M_\odot$ & km\,s$^{-1}$ \\
         \hline
         D & $36{\pm}3$ & $42{\pm}5$ & $35{\pm}8$ & $90{\pm}5$ & $11.1{\pm}1.9$ & - \\
         O   & $115{\pm}3$ & $20{\pm}4$ & $20{\pm}4$ & $120{\pm}14$ & $0.70{\pm}0.05$ & $22{\pm}4$ \\
    \hline
    \end{tabular}
    \label{tab:BFMod}
\end{table}

\section{An Outflowing Gas Interpretation}
\label{sec:modelling2}
\begin{figure*}
    \includegraphics[width=2.1\columnwidth]{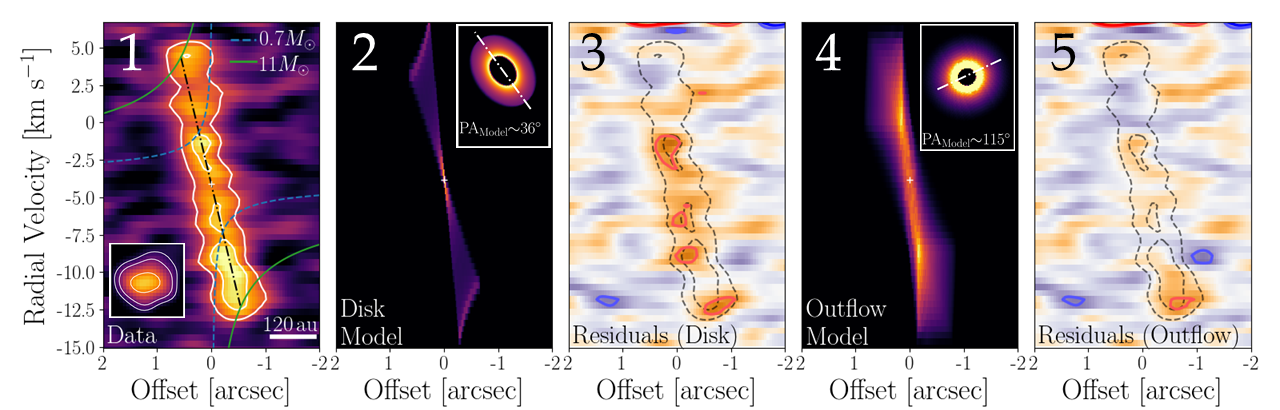}
    \caption{Plot 1: PV diagram showing the distribution of velocities present at different offsets along a slit width of 2$\arcsec$ and position angle of 36$^{\circ}$. The contours show 3 and $5\sigma$ emission. The curves demonstrate the maximum allowed radial velocities for Keplerian motion around a $11M_\odot$ star (green solid) and a $0.7M_\odot$ star (blue dashed), and the radial velocity expected for a disc with an 80\,au radius (black dot-dash slope). The lower-left thumbnail shows the Moment-0 map (box extent ${\sim}300$\,au, with 3, 6, and 9${\sigma}$ contours). Plots 2 and 4: Model PV diagrams using wider slits covering the full model extent, however we note that the emission beyond $2\arcsec$ is negligible. The thumbnails show the model Moment-0 maps (${\sim}300$\,au box widths), and PAs (see Table~\ref{tab:BFMod}). Plots 3 and 5: Residual PV diagrams, with $\pm2$ and $\pm3\sigma$ residuals shown in blue/orange, and the same contours as the data (black dashed).}
    \label{fig:NOLupPVDiag}
\end{figure*}

Having ruled out the Keplerian disc interpretation, we explore the possibility that the gas emission is dominated by a radially outflowing velocity component. 
We define an \textit{Outflow Model} with a stellar mass fixed at $0.7M_\odot$ (which sets the azimuthal velocity), where the CO emission extends between $r_{\rm{in}}$ and $r_{\rm{out}}$ with a radial velocity, $v_r$, which models the CO gas outflow. 
This is a variant of the Disc Model of $\S$\ref{sec:furtheranalysis}, with the same disc parameters and geometry, except that the gas also has an additional radial velocity component and the vertical density distribution is modelled as uniform, with fixed lower and upper edges, such that a vertical cross-section through the disc looks like a wedge with an opening angle $\delta h$. 
Allowing for a larger scale height is a rough approximation to disc wind models, where material flows both vertically off the disc and radially outwards. 
We fitted the visibilities as described above for the Disc Model. 
We again find reasonable results (see Fig.~\ref{fig:NOLupPVDiag}, plots 4 and 5), and quote the best-fit model parameters in Table~\ref{tab:BFMod}, also finding $v_{\rm{sys}}{=}-3.8\rm{km}\,\rm{s}^{-1}$, and an unconstrained $p$, which we therefore fixed as $p{=}-1$\footnote{$p{=}-1$ is the value expected for mass conservation if the CO is being created at the inner disc edge.}. 
The Outflow Model finds $v_r{\approx}22$\,km\,s$^{-1}$, which is higher than the measured velocity from the CO line width 
and thus dominates the azimuthal velocity for the modelled $r_{\rm{in}}$ and $r_{\rm{out}}$. 
The Outflow Model finds a large opening angle of ${\delta}h{=}0.3{\pm}0.1$; the disc inclination is required to be low to reproduce the spatial extent, thus the scale height is large to maximise the radial component of the velocity, and $v_r$ is significantly larger than the minimum possible value of ${\sim}19/2$\,km\,s$^{-1}$. 
Though the flow in the model is mostly radial, the large scale height can be thought of as approximating the significant vertical component present in outflow models. 
Despite the simplicity of assuming the gas to have an additional radial velocity component, this model shows that the observations are broadly consistent with a scenario in which the gas is radially outflowing. 

\section{First Imaged Class~III Gas Dispersal}
\label{sec:discussion}
Whilst we model the CO emission over all the channels in which it is detected, for simplicity we have plotted the two models of $\S$\ref{sec:furtheranalysis} and \ref{sec:modelling2} in Fig.~\ref{fig:NOLupPVDiag} as PV diagrams, with their corresponding moment-0 maps as thumbnails. 
These show that whereas the models have similar PV diagrams, the position angles of their disc midplanes are ${\sim}90^{\circ}$ different, discussed further in Appendix~\ref{sec:appendixPVDiag}. 
At low spatial resolution, Keplerian disc models and high radial velocity outflowing models (i.e., those in which the radial velocity dominates the azimuthal velocity) can have indistinguishable PV diagrams. 
This degeneracy can be broken with higher resolution imaging to measure the PA of the continuum emission which would confirm the radial (or azimuthal) nature of observed CO velocities. \\
\newline
The simple Outflow Model presented in $\S$\ref{sec:modelling2} showed that the observations can be fitted with a constant ${\sim}22$\,km\,s$^{-1}$ radial velocity component. 
While it may be possible to explain the observations with a lower outflow velocity, for example with a more detailed model of the gas kinematics, any gas flow must be ${\gtrapprox}10$\,km\,s$^{-1}$, given the spectrum line width of ${\sim}19$\,km\,s$^{-1}$. 
This is high compared to the photoevaporative models of \citet{Haworth20}, in which the outflowing wind velocity is ${\sim}3$\,km\,s$^{-1}$. 
In fact, this high velocity rules out a scenario in which the gas is a pure photoevaporative wind, since such a wind velocity should be set by the sound speed, and so to reach even ${\sim}10$\,km\,s$^{-1}$ would require the CO gas to be at $T{\sim}10,000$\,K, which exceeds the CO thermal dissociation temperature. 
To accelerate CO gas to such high speeds, additional forces are required \citep[e.g., such as MHD driven winds, see][and references therein]{Lesur20}. 
Winds thought to be driven by the disc magnetic field have been observed towards less evolved class~II stars with similarly high velocities \citep[see][]{Pontoppidan11}. 
Theoretically, magnetically driven disc winds can be launched with velocities ${\sim}$several factors of the Keplerian velocity at 10s of au \citep{Bai16}, which at the location of our disc inner edge ($r_{\rm{in}}{=}20{\pm}4$\,au) is consistent with our modelled outflow velocity. 
Thus the 10s of km\,s$^{-1}$ velocities inferred here are plausible via this mechanism.  
Though we are not aware of theoretical works that have looked into the launching of a magnetised disc wind in the conditions of a debris disc, we note that 4-12\,km\,s$^{-1}$ dust outflows were detected towards AU~Mic by \citet{Boccaletti18}, and it is worth exploring whether these are launched by a similar mechanism. 
Interpretation of NO~Lup's CO emission may therefore require modelling a 3D velocity field and MHD driving, in addition to CO photodissociation and shielding \citep[which we discuss below,][]{Kral19, Marino20}. \\
\newline
While we have discussed the parameter $v_r$ in the Outflow Model as an outflowing velocity, the fit to the observations is insensitive to the sign of $v_r$, thus we next explore whether the gas is more likely to be infalling or outflowing. 
To do so, we compare the observed CO mass loss rate with the upper limit on the CO accretion rate. 
Assuming $\dot{M}_{\rm{CO}}{=}M_{\rm{CO}}.v_r$/$R$, we find $\dot{M}_{\rm{CO}}{\sim}3M_\oplus$\,Myr$^{-1}$, for $M_{\rm{CO}}{\sim}5{\times}10^{-5}M_\oplus$, at $v_r{\sim}22$\,km\,s$^{-1}$, from the mean disc radius $R{=}$($r_{\rm{in}}$+$r_{\rm{out}}$)/2${\sim}70$\,au. 
Since the estimate of $M_{\rm{CO}}$ assumes it is in LTE and that the gas is optically thin, this mass loss rate is a lower limit. 
Next, by fitting a polynomial to the continuum near the H${\alpha}$ line (with the EW stated in $\S$\ref{sec:context}), we obtain a continuum-subtracted line flux of ${\sim}2.9{\times}10^{-15}$\,erg\,s$^{-1}$\,cm$^{-2}$, from which we estimate the accretion luminosity as $\log{L_{\rm{acc}}/L_\odot}{\sim}{-3.04}$, using the empirical relation between line luminosity and accretion luminosity of \citet{Alcala17}. 
With $L_{\rm{acc}}$, $M_\star$ and $L_\star$ from $\S$\ref{sec:context}, and a stellar radius of $R_\star{\sim}1.3R_\odot$ \citep{Gaia18}, we find a $3\sigma$ upper limit CO mass accretion rate of <0.1$M_{\oplus}$\,Myr$^{-1}$, for an ISM H$_2$/CO abundance ratio of 10,000 (i.e., consistent with the ratio in primordial gas). 
This is consistent with other non-accreting class~III stars \citep[see][]{Manara13}, however it is more than 30 times lower than the inferred CO mass loss rate, indicating that if primordial the CO gas cannot be \textit{inflowing}. 
While this cannot rule out inflowing gas for lower H$_2$/CO ratios (e.g., if the gas is produced in a secondary scenario), we rule out inflowing secondary gas later. However, the upper limit on the CO mass accretion rate may provide an important constraint on models for the outflow that require an inflowing component to conserve angular momentum. \\
\newline
To explore the origin of the gas, we first note that CO at 10-20\,km\,s$^{-1}$ (i.e., 2-4\,au\,yr$^{-1}$) would travel ${\gtrapprox}$200\,au in 100\,yr, i.e., over the CO photodissociation time \citep{Visser09}, although if well shielded the CO could survive much longer. 
This may suggest that the gas was formed during the recent break-up of a massive planetesimal, however such an event would be both rare and leave a distinctive asymmetry in the gas distribution \citep{Jackson14} which we do not observe. 
Rather, the measured ${\sim}130$\,au extent of the CO gas could still be consistent with a $>$100\,yr lifetime, as at larger distances from the star cooler CO is less collisionally excited, and so difficult to detect \citep{Matra15}. 
Thus, comparing the extent, velocity and gas lifetime does not lead to strong constraints, though this may not be the case for neutral CI gas, which is likely a better probe of outflowing gas at larger distances, as suggested by \citet{Haworth20}. 
Given typical stellar/disc timescales of 0.1-1\,Myr, the high velocity however suggests that the CO must be continuously replenished, as it is extremely unlikely that the CO was imaged within 100\,yr after a single or final production event. \\
\newline
The CO reservoir that replenishes the gas may either be in gaseous form (a protoplanetary disc remnant) or in solid form (icy planetesimals in a debris disc). 
Protoplanetary discs are expected to disperse on ${\sim}$100\,kyr timescales \citep[e.g.][]{Ercolano17}. 
While NO~Lup's SED suggests that it has already lost its protoplanetary disc, the dust may have dispersed first, leaving a primordial CO reservoir \citep[e.g., as suggested by][]{Owen19}. 
The detection of CO gas towards 1/30 of the 2\,Myr old Lupus class~III stars in \citet{Lovell20} implies plausible dispersal timescales for such primordial gas remnants of ${\sim}$70\,kyr. 
A potential problem with this, however, is that no CO reservoir with a Keplerian disc signature is present in our observations. 
For such a CO reservoir to go undetected its surface brightness, which in the optically thick limit is $I_{\nu}{=}B_{\nu}(T)A/A_{\rm beam}$ (where $A$ is the CO emission area, and $A_{\rm beam}$ is the beam area), should be below the 3$\sigma$ noise level of ${\sim}9$\,mJy\,beam$^{-1}$ \citep[see][]{Lovell20}. 
If CO emission fills the beam ($A {=} A_{\rm beam}$), then this upper limit implies a temperature below ${\sim}$4\,K, significantly below the CO sublimation temperature and thus unlikely. 
For a more reasonable temperature of 50\,K, the CO emitting area would need to be ${\sim}$150 times smaller than the beam, and thus a ring at $r_{\rm{in}}{\sim}20$\,au must have a width narrower than 0.01\,au, which is also unlikely. 
This argues against an optically thick ring of CO being the gas source, though further high-resolution imaging is required to definitively conclude this. 
A more plausible explanation may be that the wind is replenished by a reservoir of CO in icy planetesimals in the $R_{\rm{BB}}{\sim}3$\,au belt. Since blackbody disc radii estimates can be ${\gtrapprox}5{\times}$ smaller than physical planetesimal belt radii \citep[see Eq.8 of][]{Pawellek15}, this $R_{\rm{BB}}$ is consistent with the modelled inner edge of the gas. Moreover, if the CO gas in this scenario is produced in the known planetesimal belt and observed at ${\sim}130$\,au, this backs up the previous claim that the gas must be outflowing. Thus, we may be witnessing a short-lived phase \textit{after} protoplanetary disc dispersal in which CO ice is released and dispersed. 
For example, it may be that following primordial gas dispersal a previously stable reservoir of CO ice became susceptible to sublimation \citep[similar to the mechanism suggested on 486958~Arrokoth by][]{Steckloff20}, or CO was released as planetesimals were ground down in collisions \citep{Marino20}. 
If this is the case, then we may find more examples of class~III stars with rapidly dispersing gas winds. 
Gas winds have not been seen towards older gaseous debris discs which may indicate that these winds are linked to the evolutionary stage (or spectral type) of the star, or perhaps suppressed by the build-up of other gaseous species over many Myr.

\section{Conclusions}
\label{sec:conclusions}
By analysing the CO gas detected towards the class~III star, NO~Lup, we have demonstrated that the CO has a high velocity width and is spatially resolved. 
Although we showed that this can be fitted with a Keplerian Disc Model, this requires the stellar mass of NO~Lup to be implausibly high, i.e., 10 times higher than expected. 
Instead, we have shown that the gas may be outflowing with a high radial velocity, explaining the ${\sim}19$\,km\,s$^{-1}$ width and ${\sim}$130\,au spatial scale. 
We conclude that this gas is outflowing and is being continually replenished, and suggest possible interpretations. 
We note further work to explore the nature of this detection, such as new high resolution imaging of the continuum, measurements for the neutral CI gas, and detailed modelling.

\section*{Acknowledgements}
We thank the anonymous reviewer for their comments which improved the quality of this work. 
JBL is supported by an STFC postgraduate studentship. 
GMK is supported by the Royal Society as a Royal Society University Research Fellow. 
SM is supported by a Research Fellowship from Jesus College, Cambridge. 
GR acknowledges support from the Netherlands Organisation for Scientific Research (NWO, program number 016.Veni.192.233). 
M.T. has been supported by the UK Science and Technology research Council (STFC), and by the European Union's Horizon 2020 research and innovation programme under the Marie Sklodowska-Curie grant agreement No. 823823 (RISE DUSTBUSTERS project). 
JPW acknowledges support from NSF grant AST-1907486. 
MK gratefully acknowledges funding by the University of Tartu ASTRA project 2014-2020.4.01.16-0029 KOMEET. 
This work was partly supported by the Deutsche Forschungs-Gemeinschaft (DFG, German Research Foundation) - Ref no. FOR 2634/1 TE 1024/1-1. 
This work was partly supported by the Italian Ministero dell'Istruzione, Università e Ricerca (MIUR) through the grant Progetti Premiali 2012 iALMA (CUP C52I13000140001) and by the DFG cluster of excellence Origin and Structure of the Universe (www.universe-cluster.de). 

\section*{Data Availability}
This work makes use of the following ALMA data: $ADS/$ $JAO.ALMA$ $\#2018.1.00437.S$. ALMA is a partnership of ESO (representing its member states), NSF (USA) and NINS (Japan), together with NRC (Canada), MOST and ASIAA (Taiwan), and KASI (Republic of Korea), in cooperation with the Republic of Chile. The Joint ALMA Observatory is operated by ESO, AUI/NRAO and NAOJ. Based on observations collected at the European Southern Observatory under ESO programme 093.C-0506(A).


\bibliographystyle{mnras}
\bibliography{example} 


\appendix
\section{PV Diagrams}
\label{sec:appendixPVDiag}
\begin{figure}
    \includegraphics[width=1.0\columnwidth]{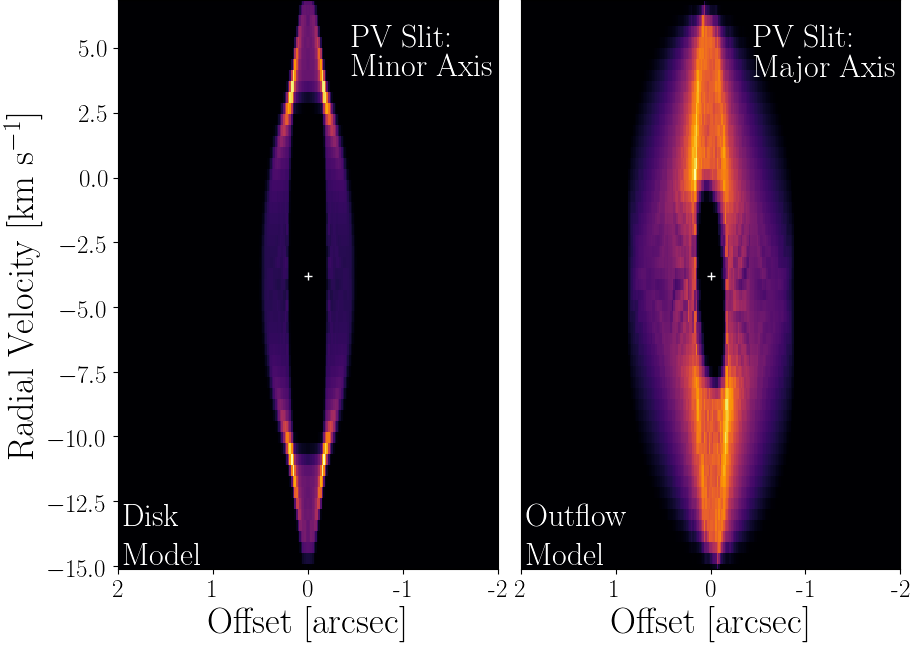}
    \caption{PV diagrams for the Disc Model and Outflow Model with their slits taken at $90^{\circ}$ to those in Fig.~\ref{fig:NOLupPVDiag} to demonstrate their elliptical nature in this orientation. The Outflow Model is rotated slightly anti-clockwise by the Keplerian motion, though this would be much greater if the radial and Keplerian velocities were more similar.}
    \label{fig:PV90}
\end{figure}

In Table~\ref{tab:BFMod} we show that the two models that fit the observations have disc position angles that are ${\sim}90^{\circ}$ different. 
Since our modelling fits all channel maps in the data set (i.e., it does not fit a PV diagram) we have a choice in how we present this. 
In Fig.~\ref{fig:NOLupPVDiag} we used a slit angle of $36^{\circ}$, i.e., parallel to direction of motion of the peak in the channel maps, the Disc Model major axis, and the Outflow Model minor axis, which produced a stripe (see panels 1, 2 and 4). 
If instead we place the PV diagram slit PA along the major axis of the Outflow Model and the minor axis of the Disc Model, these yield model PV diagrams that are then elliptical as can be seen in Fig.~\ref{fig:PV90}, and similar to those presented in \citet{Haworth20}. 
Thus, although the choice of slit PA does not bias our modelling, a different choice will result in a different visualisation, but not one that can distinguish between the Keplerian Disc Model and the Outflow Model.

\bsp	
\label{lastpage}
\end{document}